\documentclass[aps,prb,preprint,groupedaddress,superscriptaddress,longbibliography]{revtex4-1}
\usepackage[utf8]{inputenc}
\usepackage[T1]{fontenc}
\DeclareUnicodeCharacter{00A0}{~} 
\DeclareUnicodeCharacter{2212}{-} 

\usepackage{graphicx} 
\usepackage{dcolumn} 
\usepackage{bm} 
\usepackage{multirow}
\usepackage{color}
\usepackage{ulem}
\usepackage{xr}
\usepackage{cleveref}

\begin{document}

\title{Improved Electrical Conductivity of Copper and Nitrogen Functionalized Carbon Nanotubes}
\author{Mina Yoon}
\affiliation{Materials Science and Technology Division, Oak Ridge National Laboratory, Oak Ridge, Tennessee 37831, USA}
\email{myoon@ornl.gov}

\author{German D. Samolyuk}
\affiliation{Materials Science and Technology Division, Oak Ridge National Laboratory, Oak Ridge, Tennessee 37831, USA}

\author{Kai Li}
\affiliation{Buildings and Transportation Science Division, Oak Ridge National Laboratory, Oak Ridge, Tennessee 37831, USA}

\author{James A. Haynes}
\affiliation{Materials Science and Technology Division, Oak Ridge National Laboratory, Oak Ridge, Tennessee 37831, USA}

\author{Tolga Aytug}
\affiliation{Chemical Sciences Division, Oak Ridge National Laboratory, Oak Ridge, Tennessee 37831, USA}
\email{aytugt@ornl.gov}

\date{\today}

\begin{abstract}
{
In this work, we investigate the electrical conductivity of carbon nanotubes (CNTs), with a particular focus on the effects of doping. Using first-principles-based approaches, we study the electronic structure, phonon dispersion, and electron-phonon scattering to understand the finite-temperature electrical transport properties in CNTs. Our study covers both prototypical metallic and semiconducting CNTs, with special emphasis on the influence of typical defects such as vacancies and the incorporation of copper or nitrogen, such as pyridinic N, pyrrolic N, graphitic N, and oxidized N. Our theoretical study shows significant improvements in the electrical conduction properties of copper-CNT composites, especially when semiconducting CNTs are functionalized with nitrogen. Doping is found to cause significant changes in the electronic density of states near the Fermi level, which affects the electrical conductivity. Calculations show that certain types of functional groups, such as N-pyrrolic, result in more than 30-fold increase in the conductivity of semiconducting CNTs compared to Cu-incorporated CNTs alone. For metallic CNTs, the conductivity is in agreement with existing experimental data, and our prediction of significant increases in conductivity with N-pyrrolic functional group is consistent with recent experimental results, demonstrating the effectiveness of doping in modifying conductivity. Our study provides valuable insight into the electronic properties of doped CNTs and contributes to the development of ultra-high conductivity CNT composites.
}
\end{abstract}
\onecolumngrid
Notice: This manuscript has been coauthored by UT-Battelle, LLC, under Contract No. DE-AC0500OR22725 with the U.S. Department of Energy. The United States Government retains and the publisher, by accepting the article for publication, acknowledges that the United States Government retains a non-exclusive, paid-up, irrevocable, world-wide license to publish or reproduce the published form of this manuscript, or allow others to do so, for the United States Government purposes. The Department of Energy will provide public access to these results of federally sponsored research in accordance with the DOE Public Access Plan (http://energy.gov/downloads/doe-public-access-plan).

\maketitle

\section{Introduction}
\label{sec:Introduction}

Electric vehicles (EVs) have emerged as a sustainable alternative to internal combustion engine vehicles, with zero exhaust emissions, reduced greenhouse gas emissions, and lower operating costs. With governments around the world setting ambitious targets for EV adoption, there has been a surge in demand for EVs. However, challenges such as range anxiety, charging infrastructure, and energy efficiency remain key barriers to widespread adoption \cite{doe2017a, Chan2017, Bonges2016, Sovacool2018}. Range anxiety, primarily due to the limited energy storage capacity of batteries and the energy efficiency of components such as electric motors and power electronics, is a major concern for potential EV buyers. Addressing this issue will require the development of advanced materials that improve energy storage and conductivity, thereby extending the range of electric vehicles. In addition, the availability and accessibility of charging stations is critical to the adoption of EVs~\cite{doe2017b}. Consumer concerns about charging times and locations underscore the need for efficient, durable conductive materials in charging systems to increase energy transfer rates, reduce charging times, and improve the user experience. Improving the efficiency of components such as electric motors, power electronics and thermal management systems is critical to reducing energy consumption and extending the range of electric vehicles. Conductive materials with high electrical and thermal conductivity are essential.

In this context, carbon nanotubes (CNTs) offer a promising solution. Known for their exceptional mechanical, electrical and thermal properties, CNTs have attracted considerable interest for their potential applications in electronics, energy storage and nanodevices~\cite{zhou21}. Their remarkable properties make them well suited for various applications in EV technology, such as anodes and cathodes in lithium-ion batteries, improving energy storage capacity and charge/discharge rates~\cite{zhou18}. CNTs also provide mechanical stability and extended battery life by mitigating volume expansion effects during charge and discharge cycles~\cite{barzegar13}. In addition, their integration into electric motor windings can improve electrical conductivity and heat dissipation, resulting in higher power density and efficiency under varying operating conditions~\cite{raminosoa2019}. By facilitating efficient heat dissipation in power electronics, CNTs can significantly improve component reliability and performance. In addition, incorporating CNTs into composite materials can increase mechanical strength while reducing weight, leading to higher energy efficiency and longer EV range~\cite{li2020}.

The field of nanotechnology and materials science has identified CNTs as outstanding materials due to their exceptional electrical, thermal, and mechanical properties~\cite{Iijima1991,Tans1998,Charlier2007}. 
Their potential to revolutionize various technological sectors depends heavily on their electrical conductivity, a critical factor for integration into advanced electronic components and energy devices~\cite{Dresselhaus2001, Avouris2002, Baughman2002, Jang2020, Xiao2022, Tang2018}. However, achieving the desired level of conductivity in CNTs is a challenging endeavor that requires innovative approaches to manipulate and enhance this intrinsic property. Doping, a strategy borrowed from semiconductor technology, has emerged as a key method for modifying the electronic properties of CNTs~\cite{barzegar13, wang12}.  Introducing impurities into CNTs can significantly alter their electrical behavior, potentially leading to enhanced conductivity. This is critical for tailoring CNT properties to meet the specific requirements of various applications.
Despite considerable progress in understanding the electrical properties of CNTs and the influence of various doping elements, there remains a gap in the comprehensive understanding of the mechanisms by which doping affects electrical conductivity, particularly in different CNT types such as metallic and semiconducting. 

This research aims to fill this knowledge gap by investigating the effects of doping on the electrical conductivity of CNTs. Using first-principles theoretical approaches, a detailed analysis of the influence of different dopants, and the use of advanced computational methods, this study seeks to elucidate the complex interplay between doping and conductivity in both metallic and semiconducting CNTs. The knowledge gained from this research has the potential to spearhead the development of ultraconducting CNT composites, leading to the development of more efficient and powerful electronic devices and energy systems.

\section{Theoretical approaches}
\label{sec:Methodology}
We used first-principles approaches to investigate the electrical conductivity of carbon nanotubes, with particular emphasis on the influence of doping. Our research took advantage of the functionality of the PERTURBO~\cite{zhou21} package, using information on electronic structure, phonon dispersion, and electron-phonon scattering derived from density functional theory (DFT)~\cite{hohenberg64, kohn65} and density functional perturbation theory (DFPT)~\cite{{zein84, baroni01}}, in conjunction with the Quantum Espresso (QE)~\cite{giannozzi09} code. The Wannier interpolation method~\cite{{marzari12}} was applied to calculate electronic energies and velocities on an ultra-fine grid in reciprocal space using the Wanner90 code~\cite{mostofi14, pizzi20}.  The calculated electronic velocities and electron-phonon scattering rates are used to calculate electronic transport properties within the relaxation time approximation (RTZ) solution of the linearized semiclassical Boltzmann transport equation (BTE)~\cite{zhou18}. 

The study included both metallic, (5,5) and (10,10), and (11,0) semiconducting carbon nanotubes with defects such as vacancies and dopants such as nitrogen (N) or copper (Cu) atoms. The QE calculations were performed with periodic boundary conditions. The (5,5) and (10,10) CNTs were modeled using the unit cell along the CNT axis, containing 10 and 20 carbon atoms, respectively.   To minimize interactions between repeated images of CNTs in the plane perpendicular to the strain direction, the (5,5) CNT was placed in a 16$\times$16$\times$2.4628~\AA$^3$ box, while the (10,10) CNT was placed in a 30$\times$30$\times$2.468~\AA$^3$ box. The supercell of (11,0) contains 88 atoms in a 30$\times$30$\times$8.534~\AA$^3$ supercell, noting that its unit cell contains 44 atoms. In these calculations, the minimal set was duplicated along the $z$ direction to increase the Cu-Cu interaction distance and to avoid the formation of negative phonon branches while modeling the Cu impurity.

Electron-ion interactions were described using ultrasoft pseudopotentials~\cite{vanderbilt90} from the PS library~\cite{dalcorso14}.
The exchange correlation energy was calculated using the Generalized Gradient Approximation (GGA) with the Perdew, Burke, and Ernzerhof (PBE) parameterization~\cite{perdew96}. QE self-consistency cycles included Brillouin zone (BZ) summations over 1$\times$1$\times$10, 1$\times$1$\times$8, and 1$\times$1$\times$6 meshes for (5,5), (10,10), and (11,0) CNTs, respectively.
For pristine CNTs or those with vacancies, the band structure was characterized by Wannier functions representing atom-centered $p_z$ orbitals and $s$ orbitals at the center of nearest neighbor C-C bonds. The presence of N or Cu impurities in semiconducting (11,0) CNTs led to the formation of impurity characteristic bands, either within the gap for N impurities or near the gap for Cu impurities. Since these bands contribute significantly to the electrical conductivity, the focus has been on wannierization and disentanglement of these bands in systems with impurities.  For N impurities, atom-centered $s$, $p_z$, and $p_y$ orbitals were used, while for Cu impurities, $s$, $p_z$, $p_y$, and $d_{yz}$ orbitals at specific impurity atom positions were considered.

This developed parameterization facilitated the accurate calculation of electronic velocities and electron-phonon scattering rates for the calculation of electronic transport properties. The electron-phonon scattering rates were calculated on a mesh of 1$\times$1$\times$240 electron k-points and 1 $\times$1$\times$120 phonon q-points. A 0.5~meV phonon energy threshold and a 50 meV smearing of the Dirac delta function were applied. In addition, a Cauchy distribution sampling of random q-points with a Cauchy function width of 0.035 and 10000 random q-points was used in the electron-phonon self-energy calculations.

\section{Results and Discussion}
%
\begin{figure*}[h]
\centering
\includegraphics[width=6in]{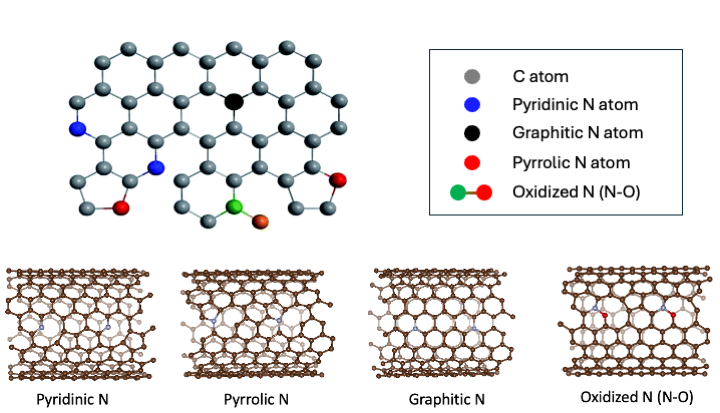}
\caption{Atomistic modeling of nitrogen-doped carbon nanotubes such as pyridinic N, pyrrolic N, graphitic N, and oxidized N. 
}
\label{fig:model}
\end{figure*}
%
This study focused on doped CNTs with significantly enhanced electrical conduction properties, supported by atomistic first-principles calculations. The energetic, electronic, and transport properties of nitrogen-functionalized CNT composites have been investigated. Our atomistic modeling of nitrogen-doped CNTs is shown in Figure~\ref{fig:model}.
Pyridinic N refers to nitrogen atoms bonded to two carbon atoms at the edge of the graphene layer, contributing $\pi$ electrons to the conjugated system. This type of nitrogen is found in the six-membered ring structure and typically enhances the basicity and electron-donating properties of CNTs \cite{Qu2010}. Pyrrolic N represents nitrogen atoms that are integrated into a five-membered ring structure, bonded to two carbon atoms, and contribute two p-electrons to the conjugated system. Pyrrolic N is known to enhance the catalytic activity and electron transfer capabilities of CNTs \cite{Liu2013}.
Graphitic N (also known as quaternary N) refers to nitrogen atoms that replace carbon atoms within the graphene plane. These nitrogen atoms are bonded to three carbon atoms and contribute to the delocalized $\pi$ system. Graphitic N can improve the electrical conductivity and stability of CNTs \cite{Tang2014}.
N-oxides in CNTs refer to a nitroso (NO) group in this study. These nitrogen oxides can introduce defects and active sites on the CNT surface, thereby enhancing their chemical reactivity and adsorption properties \cite{Wang2018}.

%
\begin{figure*}[h]
\centering
\includegraphics[width=6in]{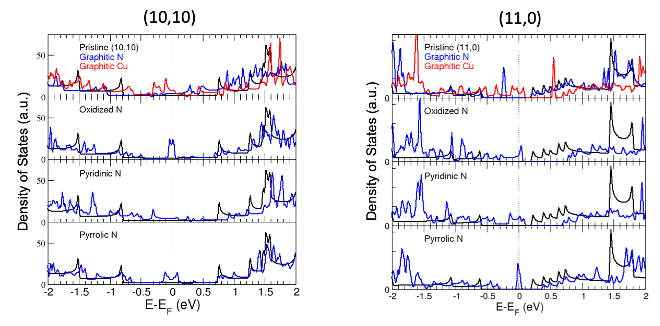}
\caption{Density of states of (10,10) metallic and (11,0) semiconducting CNTs with Cu or different N dopants, such as graphitic N, oxidized N, Pyridinic N, and Pyrrolic N. 
}
\label{fig:dos}
\end{figure*}
%
The electronic density of states (DOS) near the Fermi level showed drastic changes depending on variables such as the type of N/Cu dopants and the nature of the CNTs (metallic or semiconducting). 
Figure~\ref{fig:dos} shows that the DOS of (10,10) metallic and (11,0) semiconducting CNTs show distinct variations due to doping effects. For the metallic (10,10) CNT, doping significantly enhances the electronic density of states near the Fermi level. In particular, the incorporation of nitrogen in the form of oxidized N and pyrrolic N results in localized states with significantly increased density near the Fermi level. This increase in electronic states suggests improved electrical conductivity due to the increased availability of carriers at the Fermi level. In contrast, the semiconducting (11,0) CNT shows significant changes in DOS when doped. Doping introduces a density of states in the gap that is otherwise absent in pristine semiconducting nanotubes. In particular, the presence of pyrrolic N leads to the formation of localized states within the bandgap. These states are important because they contribute to the electronic transport properties by providing pathways for carriers to traverse the otherwise insulating gap. Consequently, pyrrolic N doping in semiconducting CNTs not only modifies the electronic structure, but also enhances conductivity by introducing states that facilitate carrier movement.

%
\begin{figure*}[h]
\centering
\includegraphics[width=3.5in]{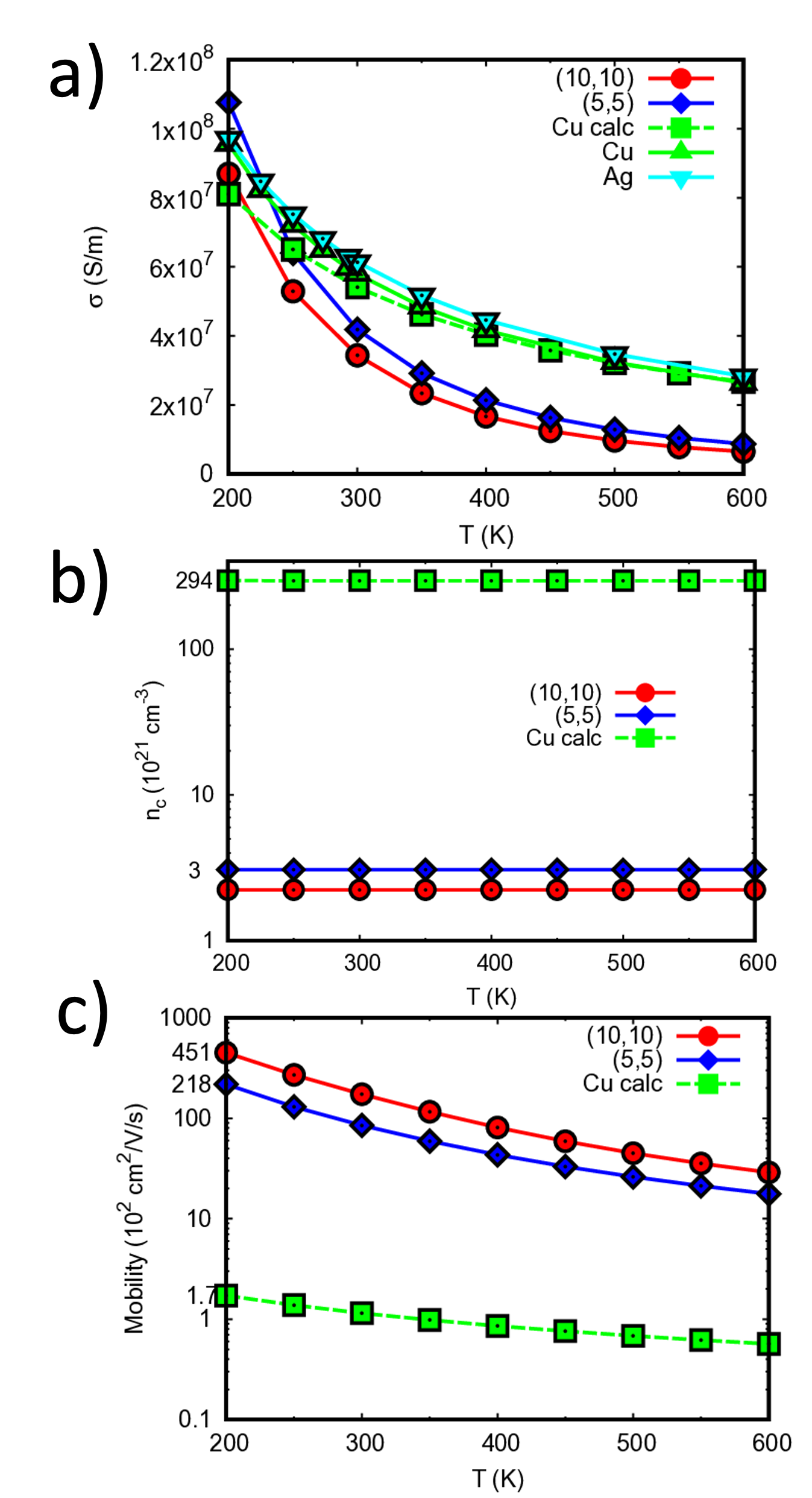}
\caption{Calculated conductivity (a), carrier density (b), and mobility (c) of (5,5) and (10, 10) CNTs and experimental data for Cu and Ag. Here the calculations are compared with the experimental data for pure bulk copper (green up-triangles), silver (cyan down-triangles), and calculation for bulk copper (green line dashed squares).
}
\label{fig:cond1}
\end{figure*}
%
Next, we evaluated the electrical conductivity of metallic and semiconducting CNTs by solving the semiclassical Boltzmann transport equation (BTE)~\cite{zhou18} (see Sec. II for details). We obtained the electrical conductivity at room temperature as high as $4.2\times10^7$ and $3.4\times10^7$~S/m in (5, 5) and (10, 10) CNTs, respectively, which is in agreement with experimental data~\cite{Miao2011} for pure CNTs, which range between $10^6$ and $10^7$~S/m~\cite{wang18}. Figure~\ref{fig:cond1} compares the calculated electrical conductivity in (5,5) and (10,10) CNTs, shown as blue diamonds and red circles, respectively, with experimental data for pure bulk copper (green up-triangles) and silver (cyan down-triangles)~\cite{matula79}.  
The calculated conductivity in (5,5) and (10,10) CNTs is somewhat smaller compared to the experimental result for Cu and Ag, i.e. $\approx 4\times10^7$ versus $\approx 6\times10^7$~S/m at 300 K. To be consistent, the calculated conductivity in CNTs is compared with the calculated conductivity in Cu and the result is shown in Fig.\ref{fig:cond1}(a) by green dashed line with squares.  The agreement between calculated (green dashed line with squares) and experimental (solid green line with triangles up) Cu conductivity is within a few percent at 300 K. This difference slightly increases with temperature decrease and starts to be almost negligible at temperature above 300 K. Such a close agreement between experiment and calculation of conductivity in Cu makes the comparison of calculated properties in CNTs and Cu relevant. 
For both (5,5) and (10,10) CNTs, the obtained conductivity at 200~K is slightly higher than the calculated one in Cu. However, with the temperature increase, the conductivity of CNTs decreases much faster than that of bulk Cu or Ag. Thus, the calculated conductivity value in Cu at room temperature is $5.4\times10^7$ S/m, while in (10,10) CNT it is $4.2\times10^7$ S/m. This difference increases with the increase of temperature - the basic understanding needs more efforts in our future study. 
To analyze the source of such difference in conductivity of CNTs and bulk Cu and Ag, we calculated the carrier concentrations and mobility.

Figure~\ref{fig:cond1}(b) presents the calculated carrier concentration (n$_c$) in (5,5), (10,10), and Cu bulk. The changes of n$_c$ with temperature are almost negligible. The carrier concentration at 200 K in Cu, $294\times10^{21}cm^{-3}$, is almost two orders of magnitude higher than in (5.5) CNT, $3\times10^{21} ~cm^{-3}$. This difference is due to the much higher density of the copper material. This significantly lower value of carrier density in CNTs compared to Cu is compensated by the higher carrier mobility in CNTs - at 200~K it is equal to $218\times10^2$ and $451\times10^2$~cm$^2$/V/s in (10,10) and (5,5) CNTs, respectively (see Fig.~\ref{fig:cond1}(d)). As a result, the conductivity at 200~K in both CNTs and Cu are close to each other (see Fig.~\ref{fig:cond1}(a)).

\begin{figure*}[h]
\centering
\includegraphics[width=6.0in]{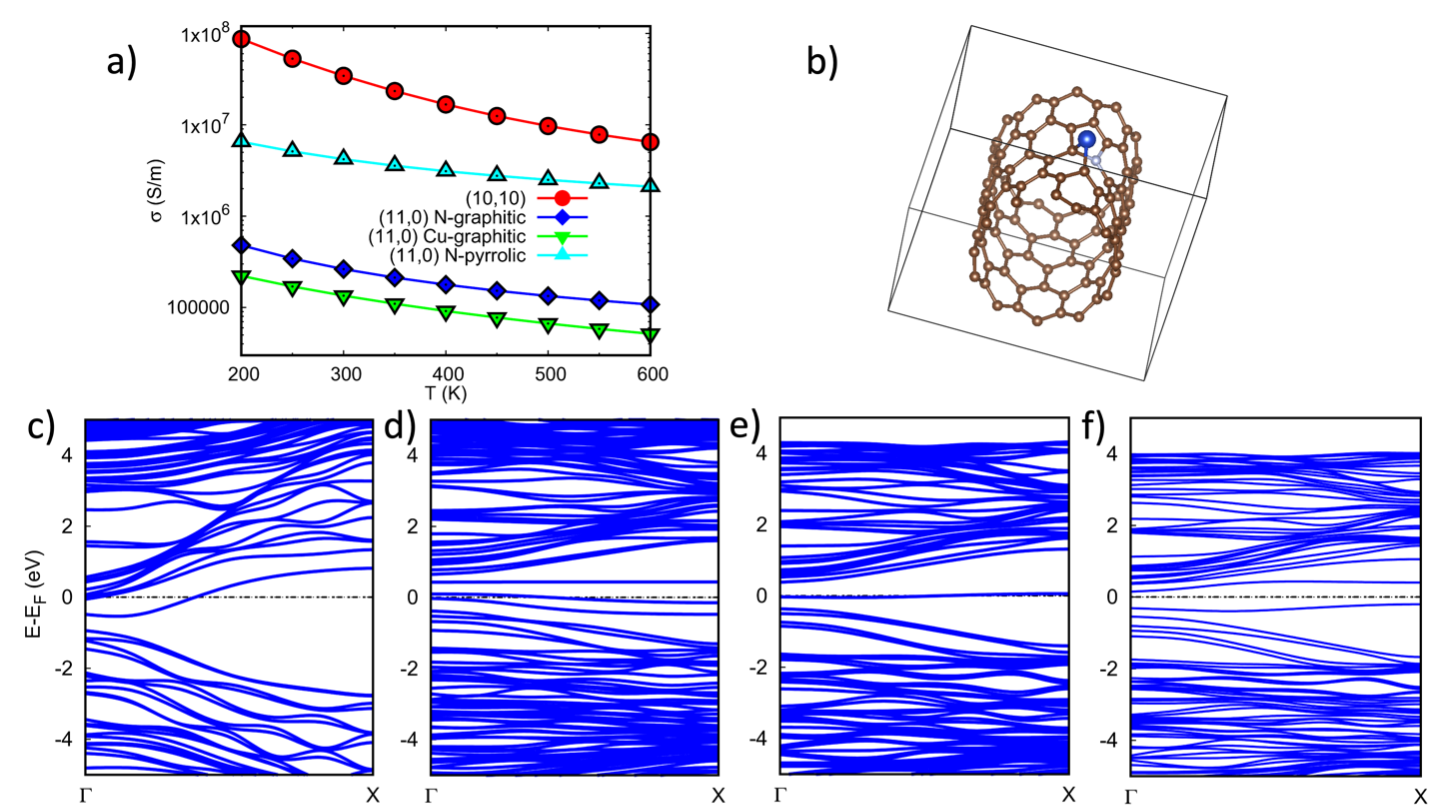}
\caption{a) calculated conductivity in (10,10) and (11,0) CNTs with N-graphitic and -pyrollic adatoms, shown by blue diamonds and cyan up triangles, respectively, and Cu-graphitic adatom, shown by green down triangles; b) N-pyrrolic plus Cu complex in (11,0) CNT structure, N atom is shown by light grey sphere, while Cu atom – by large blue sphere;  the calculated electronic structure in (11,0) CNT with c) N-, d) Cu-graphitic adatoms, e) N-pyrrolic adatom, and f) N-pyrrolic plus Cu complex shown in \ref{fig:bulk}(b).
}
\label{fig:bulk}
\end{figure*}

Semiconducting (11,0) CNTs naturally have no charge conduction at zero temperature due to a ~eV-wide gap (see SI Fig. S1), but a small number of electrons occupying the conduction bands at finite temperatures results in a minimal but non-zero conductivity. Doping, modeled with the rigid band approximation, was observed to increase this conductivity, with the carrier density enhancement increasing the conductivity of (11,0) CNTs to nearly half that of (5,5) CNTs. The effects of doping were further investigated by modeling the direct introduction of N or Cu atoms into the CNT carbon system.
It was found that to stabilize a (11,0) CNT system with a Cu-substituted carbon atom, the size of the modeled system along the direction of tube elongation must be doubled to ensure a minimum distance of 8.53~\AA between two Cu atoms. This approach avoids the appearance of negative phonon modes and consequent system instability. The introduction of N or Cu substitutions in (11,0) CNTs could lead to the formation of an additional band in the semiconducting gap, which is partially occupied by electrons (Fig.~\ref{fig:bulk}(c-f)). As a result, the room temperature conductivity of these doped CNTs increased significantly compared to their pristine counterparts, although it remained lower than that of metallic (10,10) CNTs (see Fig.~\ref{fig:bulk}(a)).

%
The introduction of dopants by replacing carbon atoms with either N or Cu can cause significant changes in the electronic band structures. A graphitic N leads to the formation of an additional band in the original gap and a shift of E$_F$ to the lower part of the conductivity electron states near the $\Gamma$ point, Fig.~\ref{fig:bulk}(c). On the other hand, a graphitic Cu dopant introduces three additional bands with $p_x$, $d_{x^2-y^2}$ and $d_{xy}$ orbital character to the CNTs, Fig.~\ref{fig:bulk}(d). These electronic structure modifications lead to changes in the conductivity character of semiconducting CNTs to metallic ones. The comparison of the calculated conductivity in the doped (11,0) CNT with the metallic (10,10) CNT is shown in Fig.~\ref{fig:bulk}(a). 
The conductivity of the (10,10) CNTs (solid red lines) is significantly larger (more than 2 orders of magnitude) than that of the doped (11,0) CNTs. At 300 K it is equal to $34.0 \times 10^6$ S/m, while in doped systems it is $0.13 \times 10^6$ and $0.26 \times 10^6$ S/m for Cu and N graphitic atoms, respectively. 

The N-pyrrolic functional group also leads to the additional partially occupied band in (11,0) CNT, Fig.~\ref{fig:bulk}(e), similar to the case of Cu or N graphitic atoms. The structure of the combination of N-pyrrolic and Cu atoms in (11,0) CNT is shown in Fig.~\ref{fig:bulk}(b), where the N atom is represented by the light gray sphere, while the Cu atoms are represented by the blue one. 
This set of bands has nitrogen $s$, $p_z$ and $p_y$ orbitals. However, the increase in conductivity is much more significant (see cyan colored lines in Fig.~\ref{fig:bulk}(a)). In contrast to the N-pyrrolic complex case, the gap is still preserved, Fig.~\ref{fig:bulk}(f), even though two additional bands are formed in the original band. As a result, the low temperature conductivity is not present. The N-pyrrolic gives the largest increase in conductivity. It is $4.2 \times 10^6$ S/m, more than 30 times higher than that of Cu-doped (11,0), and only ten times lower than that of (10,10) CNT. 
This finding is consistent with experimental evidence that nitrogen doping improves the conductivity of nanotubes with a significant proportion of semiconducting CNTs~\cite{Jiang2024}.

\section{Conclusion}
In summary, this research investigates the electrical conductivity of CNTs, focusing on the effects of doping, a key process for enhancing their electrical properties. Using first-principles approaches, the electronic structure, phonon dispersion, and electron-phonon scattering of both metallic and semiconducting CNTs are studied, with particular emphasis on understanding the effects of defects and the incorporation of nitrogen and copper atoms as dopants. The results indicate significant improvements in the electrical conduction properties of Cu-CNT composites, especially when semiconducting CNTs are nitrogen functionalized. Doping is shown to induce significant changes in the electronic density of states near the Fermi level, thereby affecting the electrical conductivity. For metallic CNTs, the conductivity obtained agrees with existing experimental data, highlighting the effectiveness of doping in modifying conductivity. For semiconducting CNTs, doping enhances conductivity, especially at room temperature, where doped CNTs exhibit significantly increased conductivity compared to their pristine counterparts, although still lower than that of metallic CNTs. Our study provides insights into these changes in conductivity, with implications for the development of ultra-high conductivity CNT composites. This advance is particularly promising for applications in EVs, addressing critical challenges such as energy storage, charging infrastructure, and overall efficiency.
\section*{Acknowledgements}
\label{sec:acknowledgments}

The research was supported by the U. S. Department of Energy, Office of Energy Efficiency and Renewable Energy, Vehicle Technologies Office, Powertrain Materials Core Program and by U.S. Department of Energy, Office of Science, Office of Basic Energy Sciences, Materials Sciences and Engineering Division. This research used resources of the Oak Ridge Leadership Computing Facility at the Oak Ridge National Laboratory, which is supported by the Office of Science of the U.S. Department of Energy under Contract No. DE-AC05-00OR22725 and resources of the National Energy Research Scientific Computing Center, a DOE Office of Science User Facility supported by the Office of Science of the U.S. Department of Energy under Contract No. DE-AC02-05CH11231 using NERSC award BES-ERCAP0024568.


\section{Reference}
\bibliography{reference.bib}

\end{document}